\documentclass{article}
\author{Manuel Arturo Izquierdo Pe\~na\footnote{{\tt maizquierdop@unal.edu.co, aizquier@cable.net.co}} \\ Observatorio Astron\'omico Nacional\\ Universidad Nacional de Colombia}
\title{The QDF file format: an electronic system to describe ancient andean khipus}
\date{May 19, 2005}

\newenvironment{example}{
    \begin{quote}\small\tt
}{
    \end{quote}
}

\newenvironment{parameters}{
    \noindent $\triangleright$ \textsf{Parameters:}
    \begin{itemize}
}{
    \end{itemize}
}

\newcommand{\qdftag}[1]{\texttt{{<#1>} {</#1>}}}
\newcommand{\qdftage}[1]{\texttt{<#1/>}}
\newcommand{\Req}{\textsf{(Required) } }
\newcommand{\Opt}{\textsf{(Optional) } }
\begin{document}
\maketitle

\begin{abstract}
With the goal of bringing to reseachers of the ancient andean khipus with a tool to share and process electronically the current corpus of these ancient information devices, I present on this paper a proposal for a Quipu Description Format \emph{QDF}, a XML based file format designed to describe such documents in a systematic and computer standard way.
\end{abstract}

\newpage
\tableofcontents
\newpage
\section {Introduction}
The fashion of the ancient andean khippus are astonishing similar to the data structures found in the modern computers. Basically, a khipu is composed of a main cord where subordinated cords with knots are attached, following a hierarchical order, very similar as the layout of directories and files present on a computer filesystem.  We can take advantage of this similarity to design a file format to describe electronically these ancient documents.

Currently the computer science brings us a very useful tool to make such description. This is the Extensible Markup Language (XML). It is a W3C-recommended general-purpose markup language for creating special-purpose markup languages. It is a simplified subset of SGML, capable of describing many different kinds of data. Its primary purpose is to facilitate the sharing of data across different systems, particularly systems connected via the Internet  (Wikipedia, 2005).

Using the principles of XML, a markup language to describe khipus can be easily designed, bringing the posibility to store those descriptions in a standard way allowing its sharing and processing easily for software packages built upon widely avaliable XML processing libraries as is LibXML2 or Expat.     

This paper will describe a proposal for such a format, denominated \textit{Quipu Description Format} (QDF), version 0.2. 

\section{QDF Structure.}
Being a format based in the XML specification, the first lines to be present in a khipu description file must be:

\begin{example}
\begin{verbatim}
<?xml version="1.0"?>
<!DOCTYPE quipu SYSTEM "qdf.dtd">
\end{verbatim}
\end{example}

These lines indicate to any XML parser the document type \texttt{(quipu)} and the Document Type Definition (DTD) wich conforms the description rules for this format. Once these lines has been put, it is mandatory to write a \texttt{<quipu> </quipu>} XML tags. These tags must enclose all the information that describes a khipu.

This information is divided in four basic sections: 

\begin{itemize}
\item \textbf{Catalog header.} 
 This section contains tags related to the identification of the khipu described. 
\item \textbf{Media index.}
This section contains indexed descriptions to features as color and materials used to build the cords of the kiphu.
\item \textbf{Metric Units.}
This small section describes the units of measure used to describe lenghts of the khipu.
\item \textbf{Maincord Description.}
This is the biggest section of a QDF file. Inside it is included the exact description of the khipu.
\end{itemize}

Each one of these sections are delimited by these tags:

\begin{center}
\begin{tabular}{ll}
Catalog header & \qdftag{about} \\
Media index & \qdftag{media\_index} \\
Metric Units & \qdftage{metric\_unit} \\
Maincord Description & \qdftag{maincord}
\end{tabular}
\end{center}

The occurence of these sections must be in the order described.

\section{Catalog Header}
This section is enclosed by the \verb+<about></about>+ tags. Inside them it must be put new tags anottating its arhaeological source, dating, reseacher code, and reseacher (who encodes that khippu) data. Allowed tags are the following (in this order):

\subsection{ \qdftag{source} }
\Req Use this tag to record the archaeological place where the khippu comes from. Also it could contain the current location, as is a museum or a private collection.

\subsection{ \qdftag{dating}  }
\Opt This tag could contain the known dating of the khipu.

\subsection{ \qdftag{codename} }
\Req There must be at last one pair of {\tt \qdftag{codename} } tags, containing the researcher code(s) used to identify the khippu. If the khipu has more than one codename, it can be put additional codename tags as necessary.

\subsection{The  \qdftag{author} subsection}
\Opt This subsection must be used to identify the data of the reseacher who make the transcription of the khippu to this QDF file. It must contain the following tags:

\subsubsection {\qdftag{name}}
\Req The name of the reseacher who generates the QDF file.

\subsubsection { \qdftag{institution}}
\Opt Institution where the reseacher belongs.

\subsubsection { \qdftag{year}}
\Opt Year of the generation of the qdf file.

\subsubsection { \qdftag{email}}
\Opt Electronic mail of the reseacher.
\subsubsection { \qdftag{address}}
\Opt Phisical Address of the reseacher.

\subsection{  \qdftag{comment}}
\Opt Use this tag to put any comment to the khippu, if necessary.

\section{Media index}
This section is enclosed by the \qdftag{media\_index} tags.
It contains a indexed list of features and characteristics of the materials used to build the khippu. These are encoded in one or several occurences of the \qdftag{material\_item} tags.

\subsection{\qdftag{material\_item}}
\Req These tags contains information required to describe both a minimal building material as a mixed material built from minimal ones already declared.

\begin{parameters}
\item {\tt label}: Brings a nickname or ID to refer that material subsequently in the file.
\end{parameters}

Inside these tags there must be the followings tags:

\subsubsection{\qdftag{description}}
\Req This tag must describe the physical nature of the medium, typical values could be ``wool'' or ``cottom'' for this tag.  In the case of a mixed material, it can hold the description of the mixing features.

\subsubsection{\qdftage{color\_rgb}}
This empty tag must describe the color of the medium, expressed with a RGB hex value.

\begin{parameters}
\item \texttt{value}: The RGB hexadecimal triplet value, in the format used as in HTML, \#rrggbb.
\end{parameters}

\subsubsection{\qdftage{color\_iccnbs}}
This tag must describe the color of the medium using a value in the ISCCNBS system.

\begin{parameters}
\item \texttt{value}: The ICCNBS value.
\end{parameters}

\subsubsection{\qdftage{mix}}
This tag is used to describe a material wich is made from the mixture of more basic materials. It is typically used when describing cords built from smaller cords of different colors and/or materials. 

\begin{parameters}
\item \texttt{id}: This refers to a a material label previously declared in the media index section.
\end{parameters}

\section{Metric units}
This section is written using only one occurence of the empty tag \qdftage{metric\_unit}. With this  small section it is specified the lenght units used in the description of the khipu.

\begin{parameters}
\item \texttt{type}: One value between \texttt{"mm"} for milimeters, \texttt{"cm"} for centimeters and \texttt{"in"} for inches.
\end{parameters}

\section{Maincord description}
This section is enclosed by the \qdftag{maincord} tags, and holds several entries of \qdftag{cord} tags, each one of these describing both pendant, top and loop cords. There could be several occurences of \qdftag{maincord} tags, when the described is a tied set of khippus. The maincord tags has also several parameters wich describe the caracteristics of the main cord of the khipu:

\begin{parameters}
\item \texttt{dir}: \Opt Describes the construction torch direction of the cord. According Urton (2003), it could have the values \texttt{"S"} or \texttt{"Z"}. If this option is not declared, a default value of \texttt{"U"} (unknown) is taken.
\item{\tt lenght}:  \Req This option shows the lenght of the maincord, measured in the units chosen in the \qdftage{metric\_unit} section.
\item{\tt width}:  \Opt It describes the width of the maincord, mesaured in the units chosen in the \qdftage{metric\_unit} section.
\item{\tt index}: \Opt  This option must be set only if the khippu described is a tied set of khippus, then each maincord tied gets a different index value. 
\item \texttt{material}: \Opt This option refers to one material label indexed in the media index section. It describes the material of the maincord.
\item \texttt{finish}: \Opt Describes how is the finishing features of the maincord. Valid values are \texttt{"knotted"} if there is an ending knot, \texttt{"broken"} if it unfortunately is broken, and \texttt{"none"} if there is no ending knot.
\end{parameters}

\subsection{Cords description.}
The \qdftag{cord} tags are the nucleus of the QDF system.
There must be a \qdftag{cord} tags for every pendant cord attached to the main cord. All the \qdftag{cord} tags needed to decribe a khipu must be inside a maincord tag. Any subsidiary cord is described as a \qdftag{cord} tags nested inside a parent \qdftag{cord} tags. This tag has the same options than the \qdftag{maincord} tags, plus some additional parameters:

\begin{parameters}
\item {\tt index}:  \Req  This index could be given in a hierarchical form, related to the main cord or the pendant this cord belong (for subsidiaries). It is advisable, by example, to use a indexing system similar to the Ascher's system (cita).

\item {\tt pos}: \Req This option must indicate the position in the parent cord measured in the units chosen in the \qdftage{metric\_unit} section, where the cord is attached. 

\item{\tt attach}:  \Opt This option indicates the attaching way of the cord to its parent cord. According to Urton (2003), it must have one of these two values: {\tt "verso"} or {\tt "recto"}. If this parameter is not given, a default value of \texttt{"U"} (unknown) is taken. 

\item \texttt{attach\_through}: \Opt This option must be set to \texttt{"yes"} if the given cord is a subsidiary that is attached through the parent cord. See the Ascher's Quipu Databook (1978) for a complete explanation of this case. If not specified, a default value \texttt{"no"} is taken.

\item {\tt type}: \Req This option indicates what kind of cord is it. Valid values are \texttt{pendant}, \texttt{top}, \texttt{subsidiary} and \texttt{loop}.

\item {\tt loop\_pos}: \Opt This option is required if this is a loop cord. Its value must indicate the position where the cord is reatached to its parent cord, measured in the units chosen in the \qdftage{metric\_unit} section.
\end{parameters}

A \qdftag{cord} item can contain four sections: Attached pendants section, media section, knots section and transcription section. 

\subsubsection{Attached pendants section}
This section is enclosed with the \qdftag{attach\_pendants} tags. This is used only in case of describing a top cord wich is attached to the maincord grouping a set of pendant cords (see Ascher,?). Inside this tag could be several occurences of the tag \qdftage{attaches}, wich posesses? this parameter:

\begin{parameters}
\item \texttt{pendant}: \Req It refers to the label given for a previous cord described.
\end{parameters}

\subsubsection{Media section}
This section is enclosed by the \qdftag{media} tags. Inside, it contains a \qdftage{material} tag, which describes what material index composes the cord. When we found cords with segments made of different materials, we can put several occurences of \qdftage{material}, according to the cord layout. The 
\qdftage{material} tag has the following parameters:

\begin{parameters}
\item \texttt{id}: \Req This option refers to one material label indexed in the media index section.
\item \texttt{pos}: \Req This is the position in the cord where that material \emph{finishes}, measured in the units chosen in the \qdftage{metric\_unit} section.
\end{parameters}

\subsubsection{Knots section}
This tags \qdftag{knots} are used to describe the knotting on the cords. Each knot must be described by a occurence of one of these tags: \qdftag{single}, \qdftag{multiple}, \qdftag{eight}, wich indentifies the known knot types. The value it sourrounds corresponds to the numerical value that it refers. These have the following parameters:

\begin{parameters}
\item \texttt{dir}: \Opt The direction of the knot. According Urton (2003), it can take one of these values \texttt{"S"} or \texttt{"Z"}. If unspecified, a default value of \texttt{"U"} (unknown) if taken.
\item \texttt{pos}: \Req The position of the knot in the cord, measured in the units chosen in the \qdftage{metric\_unit} section.
\end{parameters}

\subsubsection{Transcription section}
This optional section is enclosed by the \qdftag{transcription} tags, and holds a text wich generally describes the numerical value that is stored in the cord.

\section{QDF Example}
{\footnotesize
\begin{verbatim}
<?xml version="1.0"?>
<!DOCTYPE quipu SYSTEM "qdf.dtd">
<quipu>
   <!-- CATALOG HEADER -->
   <about>
      <source>Tocogua's Khipu</source>
      <dating>1200AD</dating>
      <codename>IZ001</codename>
      <author>
         <name>Manuel A. Izquierdo</name>
         <institution>Observatorio Astronomico. UNAL</institution>
         <year>2005</year>
         <email>maizquierdop@unal.edu.co</email>
         <address>Ciudad Universitaria, Bogota</address>
      </author>
      <comment>
         A imaginary example khipu. Based on some contents on the
         Ascher's AS132 khipu.
      </comment>
   </about>

   <!-- MEDIA INDEX -->
   <media_index>
      <material_item label="BS">
         <description>Wool ?</description>
         <color_iccnbs value="BS"/>
      </material_item>

      <material_item label="LC">
         <description>Cootom ?</description>
         <color_iccnbs value="LC"></color_iccnbs>
      </material_item>

      <material_item label="YB:LC">
         <description>Motted Wool ?</description>
         <mix id="BS"/>
         <mix id="LC"/>
      </material_item>
   </media_index>

   <!-- METRIC UNITS -->
   <metric_unit type="mm"/>
   
   
   <!-- MAINCORD DESCRIPTION -->
   <maincord material="YB:LC" lenght="600" dir="Z">
      <!-- First Cord -->
      <cord index="X1" 
         lenght="415" 
         pos="0" type="pendant" finish="knotted" dir="S">
         <media>
            <material id="LC" pos="20"/>
            <material id="YB:LC" pos="415"/>
         </media>
         <knots>
            <single pos="130">10</single>
            <single pos="132">10</single>
            <single pos="134">10</single>
         </knots>
         
         <!-- Subsidiary cord -->
         <cord index="X1s1" lenght="425" 
            pos="50" type="subsidiary" dir="Z">
            <media>
               <material id="LC" pos="425"/>
            </media>
            <knots>
               <single pos="425" dir="S">10</single>
            </knots>
            <transcription>10</transcription>
         </cord>
         <transcription>30</transcription>
      </cord>
      
      <!-- A top cord -->
      <cord index="X2" lenght="495" pos="20" type="top">
         <media>
            <material id="YB:LC" pos="501"/>
         </media>

         <!-- Top's subsidiary  -->
         <cord index="X2s1" lenght="235" pos="30" type="subsidiary">
            <media><material id="LC" pos="235"/></media>
         </cord>
      </cord>

      <!-- A loop cord -->
      <cord index="X3" lenght="501" pos="25" type="loop" loop_pos="67">
         <media>
            <material id="YB:LC" pos="501"/>
         </media>
         <!-- Loop's subsidiary -->
         <cord index="X3s1" lenght="305" pos="10" type="pendant" dir="Z">
            <media>
               <material id="YB:LC" pos="501"/>
            </media>
            <knots>
               <multiple pos="60" dir="Z">3</multiple>
               <multiple pos="110" dir="Z">7</multiple>
            </knots>
         </cord>
      </cord>
   </maincord>
</quipu>

\end{verbatim}
}
\section{Quipu Description Format DTD.}

{\footnotesize
\begin{verbatim}
<!--
Quipu Description Format DTD. Version 0.2
Copyright (C) 2005 Manuel Arturo Izquierdo P. <maizquierdop@unal.edu.co>

This library is free software; you can redistribute it and/or
modify it under the terms of the GNU Lesser General Public
License as published by the Free Software Foundation; either
version 2.1 of the License, or (at your option) any later version.

This library is distributed in the hope that it will be useful,
but WITHOUT ANY WARRANTY; without even the implied warranty of
MERCHANTABILITY or FITNESS FOR A PARTICULAR PURPOSE.  See the GNU
Lesser General Public License for more details.

You should have received a copy of the GNU Lesser General Public
License along with this library; if not, write to the Free Software
Foundation, Inc., 51 Franklin Street, Fifth Floor, Boston, MA  02110-1301  USA          
-->

<!ELEMENT quipu (about, media_index, metric_unit, maincord*)>

<!ELEMENT about (source,dating?,codename+,author?,comment?)>
    <!ELEMENT source (#PCDATA)>
    <!ELEMENT dating (#PCDATA)>
    <!ELEMENT codename (#PCDATA)>
    <!ELEMENT comment (#PCDATA)>
    <!ELEMENT author (name, institution?, year?, email?, address?)>
        <!ELEMENT name (#PCDATA)>
        <!ELEMENT institution (#PCDATA)>
        <!ELEMENT year (#PCDATA)>
        <!ELEMENT email (#PCDATA)>
        <!ELEMENT address (#PCDATA)>

<!ELEMENT media_index (material_item+)>
    <!ELEMENT material_item (description, color_rgb?, color_iccnbs?, mix*)>
        <!ATTLIST material_item label ID #REQUIRED>
        <!ELEMENT description (#PCDATA)>
        <!ELEMENT color_rgb EMPTY>
            <!ATTLIST color_rgb value CDATA #REQUIRED>
        <!ELEMENT color_iccnbs EMPTY>
            <!ATTLIST color_iccnbs value CDATA #REQUIRED>
        <!ELEMENT mix EMPTY>
            <!ATTLIST mix id IDREF #REQUIRED>
<!ELEMENT metric_unit EMPTY>
    <!ATTLIST metric_unit type (mm|cm|in) #REQUIRED>
            
<!ELEMENT maincord (cord+)>
    <!ATTLIST maincord dir (S|Z|U) #IMPLIED>
    <!ATTLIST maincord lenght CDATA #REQUIRED>
    <!ATTLIST maincord width CDATA #IMPLIED>
    <!ATTLIST maincord index ID #IMPLIED>
    <!ATTLIST maincord material IDREF #IMPLIED>
    <!ATTLIST maincord finish (knotted|broken|none) #IMPLIED>
    <!ELEMENT cord ( attach_pendants*, media, knots*, cord*,transcription?)>
        <!ATTLIST cord index ID #REQUIRED>
        <!ATTLIST cord lenght CDATA #REQUIRED>
        <!ATTLIST cord width CDATA #IMPLIED>
        <!ATTLIST cord pos CDATA #REQUIRED>
        <!ATTLIST cord dir (S|Z|U) #IMPLIED>
        <!ATTLIST cord attach (verso|recto|U) #IMPLIED>
        <!ATTLIST cord attach_through (yes|no) #IMPLIED>
        <!ATTLIST cord type (pendant|top|subsidiary|loop) #REQUIRED>
        <!ATTLIST cord loop_pos CDATA #IMPLIED>
        <!ATTLIST cord finish (knotted|broken|none) #IMPLIED>
        <!ELEMENT attach_pendants (attaches+)>
            <!ELEMENT attaches EMPTY>
                <!ATTLIST attaches pendant IDREF #REQUIRED>
        <!ELEMENT media (material*)>
            <!ELEMENT material EMPTY>
                <!ATTLIST material id IDREF #REQUIRED>
                <!ATTLIST material pos CDATA #REQUIRED>
        <!ELEMENT knots (single*, multiple*, eight*)>
            <!ELEMENT single (#PCDATA)>
                <!ATTLIST single dir (S|Z|U) #IMPLIED>
                <!ATTLIST single pos CDATA #REQUIRED>
            <!ELEMENT multiple (#PCDATA)>
                <!ATTLIST multiple dir (S|Z|U) #IMPLIED>
                <!ATTLIST multiple pos CDATA #REQUIRED>
            <!ELEMENT eight (#PCDATA)>
                <!ATTLIST eight dir (S|Z|U) #IMPLIED>
                <!ATTLIST eight pos CDATA #REQUIRED>
        <!ELEMENT transcription (#PCDATA)>

\end{verbatim}
}

\section{References}
\begin{description}
\item [ASCHER, Marcia ; ASCHER, Robert]. \\
    1978$\quad\quad$ Code of the Quipu: Databook. University of Michigan Press.
\item[URTON, Gary]. \\
    2003 $\quad\quad$ Signs of the Inka Khipu: Binary Coding in the Andean Knotted-String Records. University of Texas Press.
\item[WIKIPEDIA]. \\
    2005 $\quad\quad$ XML. (Web Resource)  \texttt{http://en.wikipedia.org/wiki/XML}
\end{description}
\end{document}